\documentclass[copyright,creativecommons,sharealike,noncommercial]{eptcs}
\providecommand{\event}{FESCA @ ETAPS 2014} 

\usepackage[english]{babel}
\usepackage[latin1]{inputenc}
\usepackage[T1]{fontenc}

\usepackage{graphicx}
\usepackage{amssymb}
\usepackage{amsmath} 
\usepackage{color}
\usepackage{url}
\usepackage{listings}
\usepackage{longtable}
\usepackage{multirow}
\usepackage{wrapfig}
\usepackage{stmaryrd}
\usepackage{colortbl}
\usepackage[lined,boxed,linesnumbered,inoutnumbered]{algorithm2e} 
\usepackage{wrapfig}
\usepackage{multicol}

\usepackage{B2}
\newcommand{\B}[1]{\lstinline[language=Bmethod,basicstyle={\sffamily \footnotesize},framesep=0pt]{#1}}

\newcommand{\siteAeLos}[0]{\scriptsize \url{http://www.lina.sciences.univ-nantes.fr/aelos/publications/fesca14}}

\definecolor{grey}{rgb}{0.95,0.95,0.95}

\newcommand{\compo}[1]{\textbf{\textsf{#1}}}
\newcommand{\reconf}[1]{\textsf{#1}}

\newcommand{\param}[1]{\textsf{#1}}

\newcommand{\AL}[1]{#1}
\newcommand{\OK}[1]{#1}

\def\p#1{\mathrel{\ooalign{\hfil$\mapstochar\mkern 5mu$\hfil\cr$#1$}}}

\def\rel{\leftrightarrow}

\def\tfun{\rightarrow}
\def\pfun{\p\tfun}
\def\tinj{\rightarrowtail}
\def\pinj{\p\tinj}
\def\tsurj{\ooalign{$\tfun$\hfil\cr$\mkern4mu\tfun$}}
\def\psurj{\p\tsurj}
\def\tbij{\ooalign{$\tinj$\hfil\cr$\mkern5mu\tfun$}}

\newcommand{\scrnp}[1]{$S_{#1} = \langle
  \mathcal{C}_{#1},\mathcal{C}^0_{#1},\mathcal{R}_{#1},\rightarrow_{#1} \rangle$}
\newcommand{\step}[1]{\stackrel{#1}{\rightarrow}} 
\newcommand{\stepr}[1]{\stackrel{#1}{\rightarrow}} 

\newtheorem{definition}{Definition}
\newtheorem{proposition}{Proposition}

\title{Component Substitution through Dynamic Reconfigurations\thanks{\OK{This work has been partially funded by the Labex ACTION, ANR-11-LABX-0001-01.}}}
\def\titlerunning{Component Substitution through Dynamic Reconfigurations}

\author{Arnaud Lanoix
\institute{LINA CNRS and Nantes University \\ Nantes, France} 
\email{arnaud.lanoix@univ-nantes.fr}
\and Olga Kouchnarenko
\institute{FEMTO-ST CNRS and University of Franche-Comt\'e \\ Besan\c{c}on, France}
\institute{Inria/Cassis France} 
\email{olga.kouchnarenko@univ-fcomte.fr}
}
\def\authorrunning{A. Lanoix \& O. Kouchnarenko}

\begin{document}
\maketitle

\begin{abstract}

Component substitution has numerous practical applications and
constitutes an active research topic. This paper proposes to enrich an
existing component-based framework---a model with dynamic
reconfigurations making the system evolve---with a new reconfiguration
operation which "substitutes" components by other components, and to
study its impact on sequences of dynamic reconfigurations.
Firstly, we define \emph{substitutability constraints} which ensure
the component encapsulation while performing reconfigurations by
component substitutions.  Then, we integrate them into a
\emph{substitutability-based simulation} to take these substituting
reconfigurations into account on sequences of dynamic
reconfigurations.  Thirdly, as this new relation being in general
undecidable for infinite-state systems, we propose a semi-algorithm to
check it on the fly. Finally, we report on experimentations using the B
tools to show the feasibility of the developed approach, and to
illustrate the paper's proposals on an example of the HTTP server.


\end{abstract}

\section{Introduction}
\label{sec:intro}

Dynamic reconfigurations~\cite{aguilar01a,nazareno02a,leger10a}
increase the availability and the reliability of component-based
systems by allowing their architecture to evolve at runtime.  In this
paper, in addition to dynamic evolution reconfigurations, possibly
guided by temporal patterns~\cite{dormoy10b,lanoix11a,dkl11:ip}, we
consider reconfigurations bringing into play by component
substitutions.  These reconfigurations by substitution may change the
model's behaviour.  The questions we are interested in are: How are
such model transformations represented? What aspects of the model's
behaviour can be changed? Can new behaviour be added, can existing
behaviours be replaced or combined with new
behaviours? 

\begin{wrapfigure}[10]{r}{.6\linewidth}
\vspace{-0.5cm}
  \centering \includegraphics[width=\linewidth]{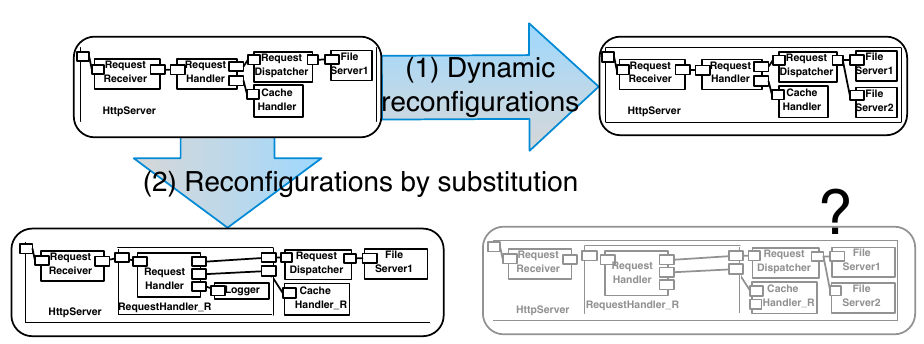}
	\caption{Different kinds of reconfigurations}
	\label{fig-principle}
\end{wrapfigure}%
More precisely, in our previous
works~\cite{dormoy10b,lanoix11a,dkl11:ip}, a component-based framework
has been developed: \OK{a component-based model} with dynamic
reconfigurations has been defined and shown consistent, a linear
temporal pattern logic allowing expressing properties over sequences
of dynamic reconfigurations has been defined. %
\OK{ In our work we suppose an \emph{interface preservation} that 
  encompasses the internal behaviour of the manipulated
  components. The approaches in~\cite{colin07a,lanoix07b} allow to deal with such an interface preservation.}

Component substitution
reconfigurations being motivated by numerous practical applications,
this paper proposes to enrich the existing component-based framework
with a notion of \emph{component substitutability}.  
Figure~\ref{fig-principle} displays two kinds of reconfigurations: \emph{Horizontal reconfigurations} represent the
dynamic architecture's evolution whereas \emph{vertical substitutions}
lead to different implementations. As the model and its
implementations must remain consistent through evolution, \OK{in this paper} we study the
impact of reconfigurations by substitution \OK{(vertical substitutions)} on sequences of dynamic
reconfigurations \OK{(horizontal reconfigurations)}.

Since our component-based \OK{model} is
formulated as a theory in \OK{first order logic (FOL)}, this is achieved by introducing a new
relation over components, and a set of logical constraints. Then, the
paper presents a notion of simulation between dynamic reconfigurable
systems wrt. a given component substitution relation, and addresses
the checking of this relation, which is known to be, in general,
undecidable.



%
%

{\em{Layout of the paper.}} In Sect.~\ref{sec:model} we recall the
main features of the architectural reconfiguration model introduced
in~\cite{dormoy10b,lanoix11a} and illustrate them on an example of the
HTTP server.  In Sect.~\ref{sec:substitutability}, a new
reconfiguration operation by component substitution is introduced and
substitutability constraints are defined to ensure component
encapsulation.  In Sect.~\ref{sec:relSubs} component substitutability
is integrated into a substitutability-based simulation relation. This
relation being undecidable in general, a semi-algorithm is proposed to
evaluate on the fly dynamic reconfiguration sequences and,
consequently, the component substitutability-based
simulation. Section~\ref{sec:experimentation} explains how to use the
B tools for dealing with component substitutability through dynamic
reconfigurations, and describes experiments on the HTTP server
example.  Finally, we conclude in Sect.~\ref{sec:conclu}.


\section{Background: Architectural Reconfiguration Model}
\label{sec:model}

The (dynamic) reconfigurations we consider here make the
component-based architecture evolve dynamically.  They are
combinations of {\em primitive} operations such as
instantiation/destruction of components; addition/removal of 
\AL{sub}components \AL{ to/from composite ones}; binding/unbinding of
component interfaces; starting/stopping components; setting parameter
values of components.  \AL{In the remaining of the paper, these
  primitive operations are not \OK{considered}, we only focus on their
  combinations \OK{providing} example-specific reconfigurations.}

\begin{wrapfigure}[9]{r}{.4\linewidth}%
\vspace{-0.6cm}
\includegraphics[width=\linewidth]{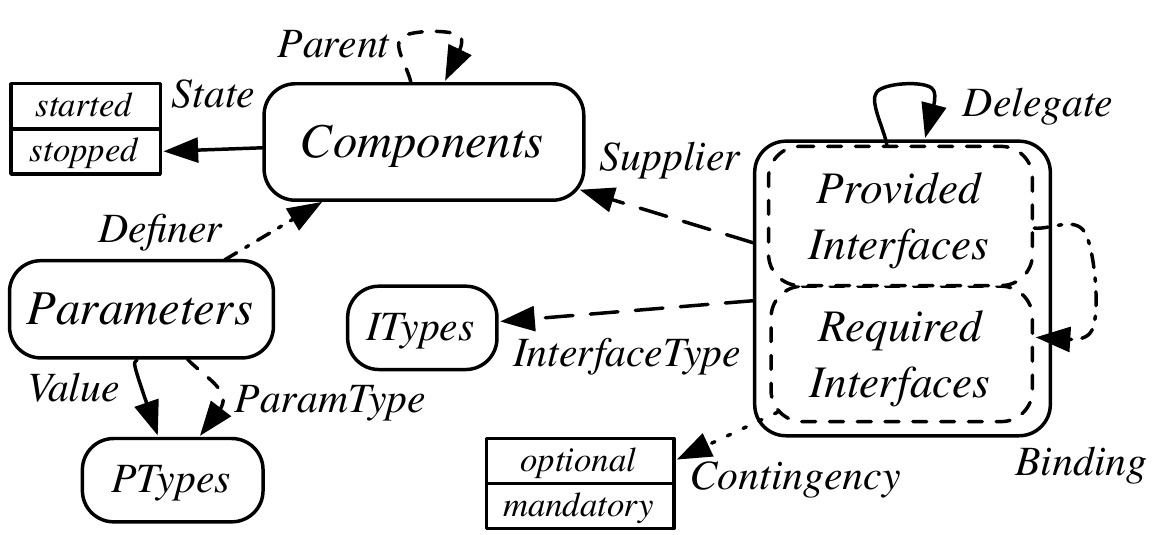}
\vspace{-0.7cm}
 	\caption{Configurations = architectural elements and relations} 
 	\label{fig-config-model}
\end{wrapfigure}
In general, system configuration is the specific definition of the
elements that define or prescribe what a system is composed of.  We
define a configuration to be a set of architectural elements
(components, required or provided interfaces and parameters) together
with relations to structure and to link them, as depicted in
Fig.~\ref{fig-config-model}~\footnote{See Definition~\ref{def:config}
  in Appendix~\ref{appendix:config}}.

Given a set of configurations $\mathcal{C} = \{c,
c_{1},c_{2},\ldots\}$, we introduce a set $CP$ of configuration
properties on the architectural elements and the relations between
them.  These properties are specified using first-order logic
formulas.  The interpretation of functions, relations, and
predicates 
is done according to basic definitions in~\cite{Hamilton78} and
in~\cite{dkl11:ip}$^1$.  We now define a configuration
\emph{interpretation} function $l : \mathcal{C} \rightarrow CP$ which
gives the largest conjunction of $cp \in CP$ evaluated to true on $c
\in \mathcal{C}$~\footnote{By definition in~\cite{Hamilton78}, this
  conjunction is in $CP$.}.

Among all the configuration properties, we consider the architectural
\emph{consistency constraints} $CC$ which express requirements on
component assembly common to all the component architectures.  They
allow defining \emph{consistent configurations} which notably respect
the following rules. Their intuition is as follows, together with a
formal description for several constraints\footnote{\AL{The whole
  definition is available at \siteAeLos.}}:
\begin{itemize}
\item a component \emph{supplies} one provided interface, at
  least;
\item the composite components do not have any parameters;
\item a sub-component must not be a composite including its own parent
  component;
%
\item two bound interfaces must have the same interface type; they are
  not supplied by the same component, but their containers are
  sub-components of the same
  composite;
    {\footnotesize $$\begin{array}{l}
        \forall ip \in ProvInterfaces, \\
        \forall ir \in ReqInterfaces
      \end{array} . 
      \left(  Binding(ip)= ir \Rightarrow 
        \begin{array}{l}
            InterfaceType(ip) = InterfaceType(ir) \\
            \wedge\ Container(ip) \neq Container(ir) \\
            \wedge\  \exists~c \in Components . 
          \left( \begin{array}{l}
              (Container(ip), c) \in Parent \\ 
              \wedge (Container(ir), c) \in Parent  \\
            \end{array} \right)  
        \end{array} \right)  
$$}
\item when binding two interfaces, there is a need to ensure that they
  have not been involved in a delegation
  yet; similarly, when establishing a
  delegation link between two interfaces, the specifier must ensure
  that they have not been involved in a binding
  yet;
\item a provided (resp. required) interface of a sub-component is
  delegated to at most one provided (resp. required) interface of its
  parent component; the interfaces involved in the delegation must
  have the same interface
  type;
\item a component is $started$ only if its mandatory required
  interfaces are bound or delegated.
\end{itemize}

\begin{definition}[Consistent configuration]
  Let $c=\langle Elem, Rel\rangle$ be a configuration and $CC$ the
  architectural consistency constraints.  The configuration $c$ is
  \emph{consistent}, written \FuncSty{consistent($c$)}, if $l(c)
  \Rightarrow CC$.
\label{def:consistent}
\end{definition}

Let $\mathcal{R}$ be a finite set of reconfiguration operations.  The
possible evolutions of the component architecture via the
reconfiguration operations are defined as a transition system over
$\mathcal{R}$.

\begin{definition}[Reconfiguration model]
  The operational semantics of component systems with reconfigurations
  is defined by the labelled transition system \scrnp{} where
  $\mathcal{C} = \{c, c_{1},c_{2},\ldots\}$ is a set of
  \emph{consistent} configurations, $\mathcal{C}^0 \subseteq
  \mathcal{C}$ is a set of initial configurations, $\mathcal{R}$ is a
  finite set of reconfigurations, $\rightarrow$ $\subseteq \mathcal{C}
  \times \mathcal{R} \times \mathcal{C}$ is the reconfiguration
  relation.
\end{definition}

\begin{wrapfigure}[6]{r}{.5\linewidth}
\vspace{-0.6cm}
	\centering
	\includegraphics[width=\linewidth]{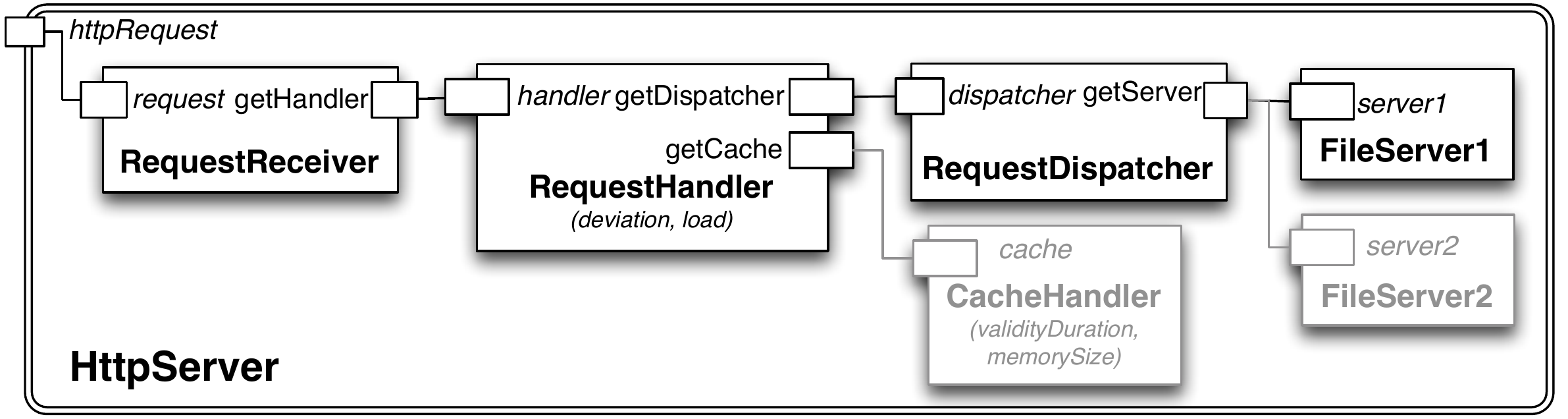}
	\vspace{-0.9cm}\caption{HTTP server architecture}
	\label{fig-http}
\end{wrapfigure}%
Let us write $c \step{ope} c'$ when a target configuration $c'$ is
reached from a configuration $c$ by a reconfiguration operation $ope
\in \mathcal{R}$.  Given the model \scrnp{}, an evolution path
$\sigma$ (or a path for short) in $S$ is a (possibly infinite)
sequence of configurations $c_0, c_{1}, c_{2}, \ldots$ such that
$\forall i\ge~0~.~(\exists\ ope_{i} \in \mathcal{R} . (c_{i}
\step{ope_{i}} c_{i+1} \in \rightarrow))$.  We write $\sigma(i)$ to
denote the $i$-th configuration of a path $\sigma$.  Let $\Sigma$
denote the set of paths, and $\Sigma^f$ ($\subseteq \Sigma$) the set
of finite paths. 

To illustrate our model, let us consider an example of a HTTP
server~\footnote{The example specification is available at
  \url{http://fractal.ow2.org/tutorial}.}.  The architecture of this
server is depicted in Fig.~\ref{fig-http}.  The
\compo{RequestReceiver} component reads HTTP requests from the network
and transmits them to the \compo{RequestHandler} component. In order
to keep the response time as short as possible, \compo{RequestHandler}
can either use a cache (with the component \compo{CacheHandler}) or
directly transmit the request to the \compo{RequestDispatcher}
component. The number of requests (\param{load}) and the percentage of
similar requests (\param{deviation}) are two parameters defined for
the \compo{RequestHandler} component.  The \compo{CacheHandler}
component is used only if the number of similar HTTP requests is
high. The \param{memorySize} for the \compo{CacheHandler} component
depends on the overall \param{load} of the server.

\begin{wrapfigure}{r}{.7\linewidth}
\vspace{-0.5cm}
 \centering
  \includegraphics[width=\linewidth]{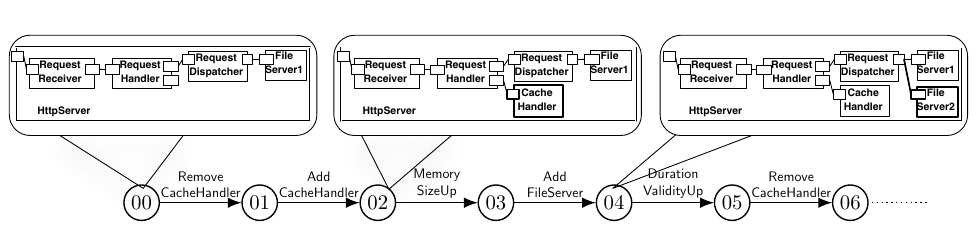}
\vspace{-0.8cm}
  \caption{Part of a path of the HTTP server architecture}
  \label{fig-trace-example}
\end{wrapfigure}%
The \param{validityDuration} of data in the cache also depends on the
overall \param{load} of the server.  The number of used file servers
(like the \compo{FileServer1} and \compo{FileServer2} components) used
by \compo{RequestDispatcher} depends on the overall \param{load} of
the server. On this example, the considered reconfiguration operations
are:
\begin{itemize}
\item \reconf{AddCacheHandler} and
  \reconf{RemoveCacheHandler} which are used to add and
  remove \compo{CacheHandler}; 
\item\reconf{AddFileServer}
  and \reconf{removeFileServer} which are used to add and
  remove \compo{FileServer2};
\item \reconf{MemorySizeUp} and
  \reconf{MemorySizeDown} which are used to increase and
  to decrease the \param{MemorySize} value;
\item \reconf{DurationValidityUp} and \reconf{DurationValidityDown}
  which are used  to
  increase and to decrease the \param{ValidityDuration}
  value. 
\end{itemize}A possible evolution path of the HTTP server architecture is
  given in Fig.~\ref{fig-trace-example}.


\section{New Reconfigurations by Component Substitution}
\label{sec:substitutability}
\begin{wrapfigure}[14]{r}{.5\linewidth}
\vspace{-0.4cm}
\centering \includegraphics[width=.95\linewidth]{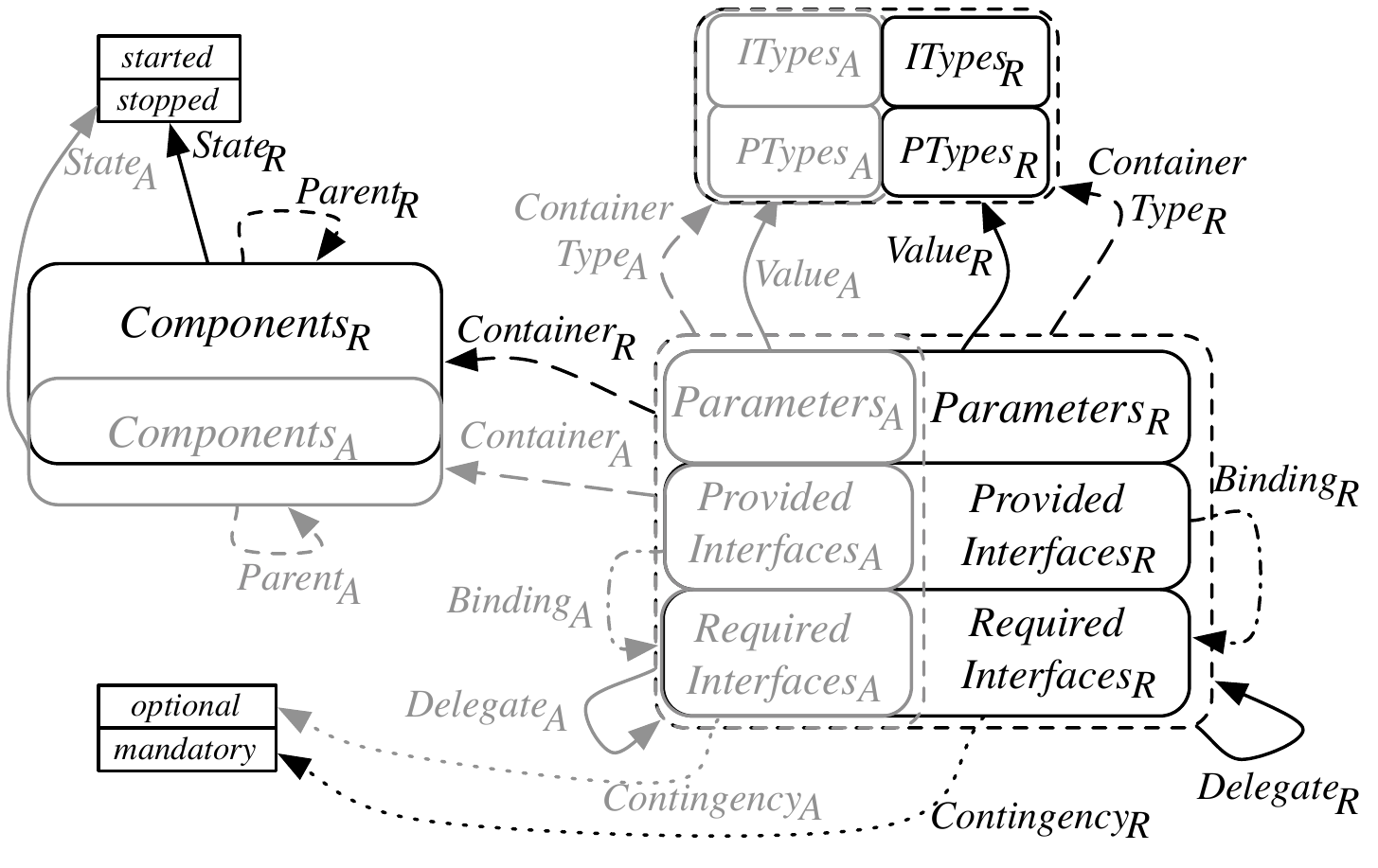}
\caption{Architectural elements before(grey) and after(black)
  substitution}
	\label{fig-config-model-raff}
\end{wrapfigure}%
In this section we enrich our component-based framework with a new
kind of reconfigurations allowing a \emph{structural} substitution of
the components with respect to the component encapsulation.  \AL{We
  suppose an \emph{interface preservation} encompassing the internal
  behaviour of the considering components, i.e. using the same interface  
  implies the same internal component behaviour~\cite{colin07a,lanoix07b}.}  We
want the substituted component to supply the same interfaces of the
same types as before. This way the other components do not see the
difference between the component and its new ``substituted'' version,
and thus there is no need to adapt them. As the substitution of a
component should not cause any changes outside of this component, only
the two following kinds of component \emph{substitutions} are allowed:
\begin{itemize}
\item either a component can be replaced by a new version of itself, or
\item a component can be replaced by a
composite component which encapsulates new sub-components providing at
least the same functionalities as before substitution.
\end{itemize}

For the allowed substitution cases, Figure~\ref{fig-config-model-raff}
displays how the architectural elements and relations are defined at
two pre- and post-substitution levels.  Let $c_A$ and $c_R$ be two
architectural configurations at respectively a pre-substitution and a
post-substitution levels.  The \emph{substitute reconfiguration} is then expressed
by a {\em partial} function $Subst\, :\,
Components_A \rightarrow Components_R$ that gives how the components are 
substituted in $c_A$ to obtain $c_R$.

\begin{wrapfigure}[12]{r}{.6\linewidth}
  \centering
  \includegraphics[width=\linewidth]{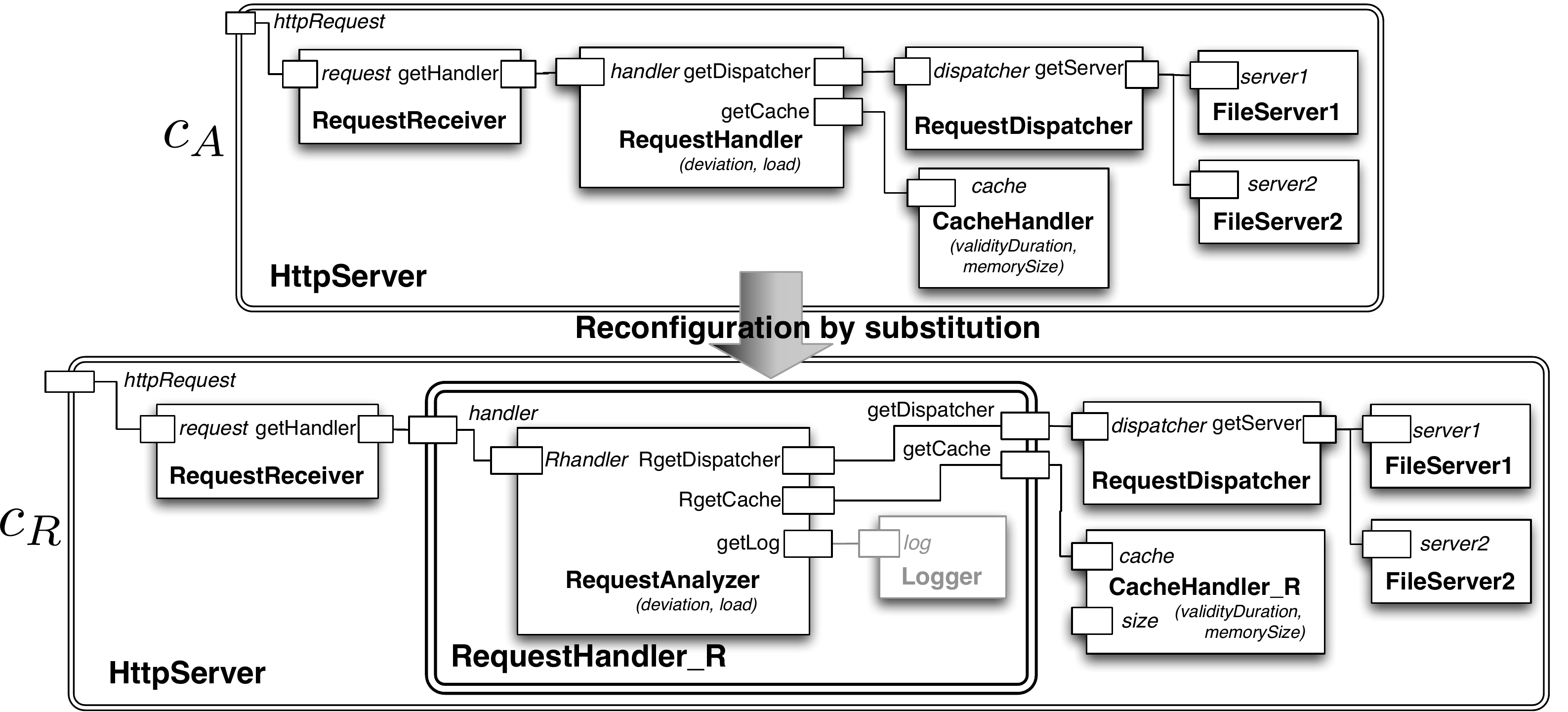}
\vspace{-0.6cm}
  \caption{Applying a reconfiguration by substitution on the \compo{HttpServer} example}
  \label{fig:httpRaff}
\end{wrapfigure}%
  Let us illustrate our proposal on the example of the HTTP server.
  For the configuration in Fig.~\ref{fig:httpRaff}, we apply the following substitute reconfiguration: 
  \begin{itemize}
  \item \compo{CacheHandler} is replaced by a new version of itself,
    named \compo{CacheHandler\_R};
  \item \compo{RequestHandler} becomes a composite component, called
    \compo{RequestHandler\_R}, which encapsulates two new
    components: \compo{RequestAnalyzer} and
    \compo{Logger}. \compo{RequestAnalyzer} handles requests to
    determine the values of the \param{deviation} and \param{load}
    parameters.  \compo{Logger} allows \compo{RequestAnalyzer} to
    memorise requests to choose either \compo{RequestDispatcher} or
    \compo{CacheHandler}, if it is available, to answer requests.
 \end{itemize}
We have {\small $ \left( \begin{array}{l}
	Subst(\text{\compo{CacheHandler}}) = \text{\compo{CacheHandler\_R}} \\
    Subst(\text{\compo{RequestHandler}}) = \text{\compo{RequestHandler\_R}}
    \end{array}\right) $} as substitute reconfiguration function.

\medskip In order to ensure that proposed substitutions respect the
requirements on components and their assembly, we now introduce
\emph{architectural constraints} on both replaced (or old) and
substituted (or new) components. These architectural constraints,
named $SC_{Subst}$, describe which changes are allowed or prescribed
by a substitute reconfiguration.  Their intuition is as follows,
together with a formal description for several
constraints\footnote{\AL{The whole definition is available at \siteAeLos.}}:
\begin{itemize}
\item \textit{In the system parts not concerned by the component
    substitution, all the core entities and all the relations between
    them remain unchanged through the substitution process:}
  \begin{itemize}
  \item the old parameters and the associated types remain unchanged
    in the substitutes;
  \item the old components remain unchanged;
{\footnotesize $$\begin{array}{r}
\forall c \in Components _A \cap Components_R,\\
\forall x \in Interfaces_A \uplus Parameters_A
\end{array}
.  \left(Container_A(x) = c \Rightarrow  Container_R(x) = c\right)$$}
  \item the old interfaces and their types are not
    changed; 
  \item the old connections between component's interfaces are kept as well.
\end{itemize}
\item \textit{For the old components impacted by the components
    substitution, the constraints are as follows:}
\begin{itemize} 
\item an old component completely disappears only if it is substituted
  by a new version for itself;
  {\footnotesize $$\forall c_A . 
\left(
\begin{array}{r}
c_A \in Components _A \\ 
\setminus Components_R 
\end{array} 
\Rightarrow \left( 
\begin{array}{r}
\exists c_R \in Components_R \\
\setminus Components_A 
\end{array} 
. \left(Subst(c_A) = c_R\right)
\right) \right)$$}
\item the substituted components are in the same state as the old
  ones, and either they have the same parent component as before
  substitution, or the old parent component has been substituted as
  well;
\item the interfaces of the replaced components are supplied by the
  substituted components;
\item the parameters of the replaced components are defined either on
  the substituted components, or on their  subcomponents.
\end{itemize}
\item \textit{The new elements introduced during the substitution
    process cannot impact the old conserved architecture:}
\begin{itemize}
\item the newly introduced components must be subcomponents of some
  substituted components;
 {\footnotesize $$
\begin{array}{r}
\forall c_R \in Components_R \setminus Components_A, \\
\forall c_A \in Components _A \setminus Components_R
\end{array} 
. \left( 
Subst(c_A) \ne c_R \Rightarrow 
\begin{array}{l}
\exists c'_R \in Components_R \setminus Components_A . \\
\phantom{aaa} \left( (c_R,c_R') \in Parent_R \right)
\end{array} \right)
$$}
\item the newly introduced interfaces must be associated with the new
  components;
  {\footnotesize $$\forall i .
\left(  i \in 
\begin{array}{l}
ProvInterfaces_R \\
 \setminus ProvInterfaces_A
\end{array}
\Rightarrow
Container_R(i) \in 
\begin{array}{l}
Components_R \\
\setminus Components_A
\end{array}\right)$$}
\item the newly introduced parameters are associated with the new
  components;
\item the new connections are used to connect the new
  components.
\end{itemize}
\end{itemize}

\begin{definition}[Structural substitutability]
    \label{def:structuralRefinement}
    Let $c_A$ and $c_R$ be two \emph{consistent} configurations, $Subst$ the
    substitution function, and $SC_{Subst}$ the architectural
    substitutability constraints.  The configuration $c_R$ is
    \emph{substitutable} to $c_A$, written \FuncSty{subst($c_R$,
      $c_A$)}, if $l(c_R) \wedge SC_{Subst} \Rightarrow l(c_A)$.
\end{definition}


\section{Component Substitution through Dynamic Evolution}
\label{sec:relSubs}
The new reconfigurations by component substitution defined in
Sect.~\ref{sec:substitutability} must be taken into account in
evolutions of component-based architectures.  Indeed, as the
substituted or the newly introduced components may introduce
\emph{new} dynamic reconfigurations, the architectures with
substituted components may evolve by the old \OK{(i.e., existing before component substitution)} reconfigurations  as well
as by new reconfigurations.  We want these \OK{(horizontal in Fig.~\ref{fig-principle})} reconfigurations to be
consistent with the reconfigurations by substitution \OK{(vertical in Fig.~\ref{fig-principle})}. To this end, we
integrate the architectural substitutability constraints from
Sect.~\ref{sec:substitutability} into a simulation relation linking
dynamic reconfigurations of a system after component's substitutions
with their old counterparts \AL{that were possible before the component
  substitution}.

\begin{figure}[htbp]
  \centering
  \includegraphics[width=\linewidth]{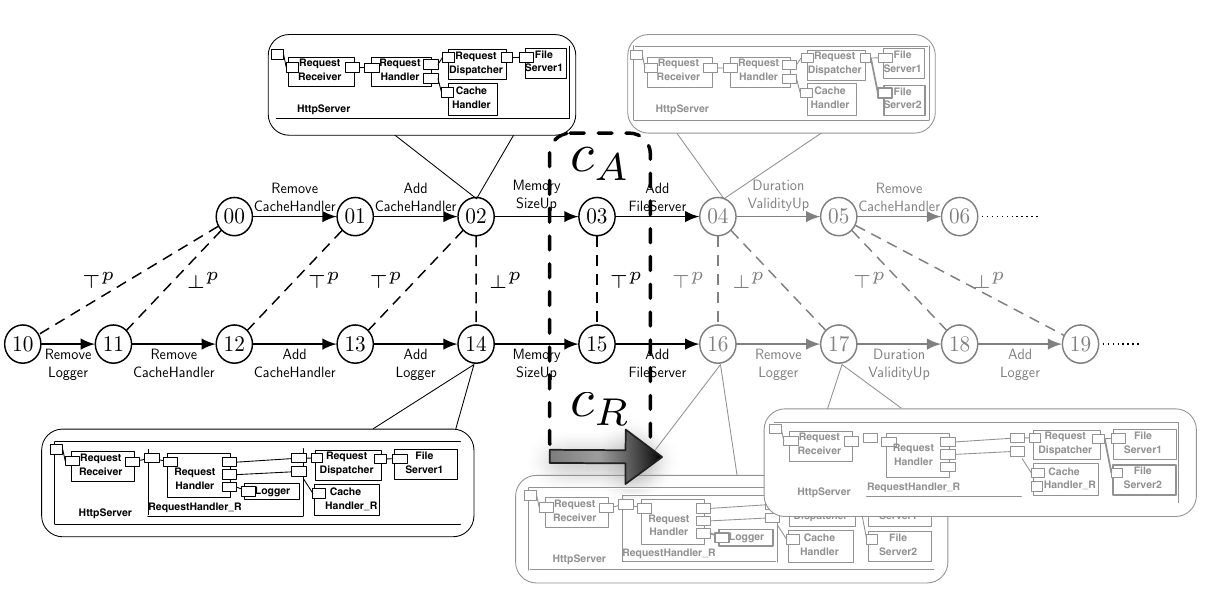}
  \caption{Substitutability evaluation at runtime}
  \label{fig:example-subs}
\end{figure}

\AL{Let us illustrate our purpose on the example displayed in
  Fig.~\ref{fig:example-subs}.  As new dynamic reconfigurations
  introduced by the component substitution, we consider
  \reconf{AddLogger} and \reconf{RemoveLogger} which consist
  respectively in adding or removing the newly introduced
  \compo{Logger} component (see Fig.~\ref{fig:httpRaff}). These new
  dynamic reconfigurations must preserve the old configurations
  sequences.}

\medskip We then define a substitution relation $\rho$ in the style of
Milner-Park~\cite{Milner:1989} as a simulation having the following
properties, which are common to other formalisms like action
systems~\cite{Butler96} or LTL refinement~\cite{KestenMP93}:
\begin{enumerate}
\item Adding the new dynamic reconfiguration actions should not introduce
  deadlocks\footnote{We write $c_R \not \rightarrow$ to mean that
    $\forall ope, c'.\ c \stepr{ope} c' \not \in \rightarrow$.}.
\item Moreover, the new dynamic reconfiguration actions should not take
  control forever: the livelocks formed by these actions are
  forbidden.
\end{enumerate}

\begin{definition}[Substitutability-based simulation]
  \label{def:relationraff}
  Let \scrnp{A} and \scrnp{R} be two reconfiguration models.
  Let $\sigma_R$ be a path of $S_{R}$.
  A relation $\sqsubseteq_{\FuncSty{subst}}
  \subseteq \mathcal{C}_R \times \mathcal{C}_A$ is the
  \emph{substitutability-based simulation} iff whenever $c_R
  \sqsubseteq_{\FuncSty{subst}} c_A$ then it implies: structural
  substitutability~\eqref{refinement:0}, strict
  simulation~\eqref{refinement:1}, stuttering
  simulation~\eqref{refinement:2}, non introduction of
  divergence~\eqref{refinement:3}, and non introduction of
  deadlocks~\eqref{refinement:4}, defined as follows: 
 \renewcommand{\theequation}{\roman{equation}} 
\begin{eqnarray}
   \FuncSty{subst$(c_R, c_A)$}  \label{refinement:0}\\
   \forall c'_R \in \mathcal{C}_R,  ope \in
    {\mathcal{R}}_R \cap {\mathcal{R}}_A . ( c_R \stepr{ope} c'_R \
       \Rightarrow\ \exists c'_A \in \mathcal{C}_A . ( c_A \stepr{ope}
       c'_A \wedge c'_R \sqsubseteq_{\FuncSty{subst}} c'_A ))
      \label{refinement:1}\\
   \forall c'_R \in \mathcal{C}_R, ope' \in
   {\mathcal{R}}_R \setminus {\mathcal{R}}_A . (
         c_R \stepr{ope'} c'_R
         \ \Rightarrow\ 
         c'_R\sqsubseteq_{\FuncSty{subst}} c_A
         ) 
        \label{refinement:2}\\
        \forall c'_R \in 
       \mathcal{C}_R , ope' \in
   {\mathcal{R}}_R \setminus {\mathcal{R}}_A, k . (
          k \ge 0 \wedge  c_R = \sigma_R(k) \wedge  c_R \stepr{ope'} c'_R
         \Rightarrow \begin{array}{l}
\exists k',  ope \in
   {\mathcal{R}}_R \cap {\mathcal{R}}_A. \\ (
          k' > k \wedge \sigma_R(k') \stepr{ope} \sigma_R(k'+1) 
          )) 
         \end{array}
\label{refinement:3}\\
\forall c_A \in \mathcal{C}_A, 
         \forall c_R \in \mathcal{C}_R . (
         c_R \sqsubseteq_{\FuncSty{subst}} c_A \wedge c_R \not \rightarrow 
         \ \Rightarrow\  
         c_A \not\rightarrow
         )
         \label{refinement:4}
   \end{eqnarray}
\end{definition}

\noindent We call the substitutability-based simulation (or the
substitutability for short) the greatest binary relation over the
configurations of $S_R$ and $S_A$ satisfying the above definition. We
say that $S_R$ \emph{is simulated by} $S_A$ wrt. the component
substitutability, written $S_R \sqsubseteq_{\FuncSty{subst}} S_A$, if
$ \forall c_R . (c_R \in \mathcal{C}^0_R \Rightarrow \exists c_A
. (c_A \in \mathcal{C}^0_A \wedge c_R \sqsubseteq_{\FuncSty{subst}}
c_A)) $.

\medskip The substitutability-based simulation defined above can be
viewed as a divergence sensitive stability respecting completed
simulation in van Glabbeek's spectrum~\cite{Glabbeek93}. Since the
models are infinite state, the problem to know whether the
substitutability-based simulation holds or not is undecidable in
general. 
Actually, as the clauses of the
substitutability relation $\sqsubseteq_{\FuncSty{subst}}$ depend not
only on the current configurations but also on the target
configurations, and even more on sequences of future configurations as
in~\eqref{refinement:3}, in general they cannot be evaluated to true
or false on the current pair of configurations. But, on the other
hand, if one of the clauses of Def.~\ref{def:relationraff} is
evaluated to \emph{false} on finite parts of the reconfiguration
sequences, then obviously the whole relation does not hold. So,
instead of considering the whole transition systems, let us consider a
sequence of reconfigurations before substitutions and its counterpart
obtained by applying reconfigurations by substitution.

\medskip We propose a semi-algorithm displayed in Fig.~\ref{fig:algo}
to evaluate on the fly the substitutability-based simulation starting
from the initial configurations $c_R^0 \in \mathcal{C}_R^0$, $c_A^0
\in \mathcal{C}_A^0$.  This semi-algorithm uses the following
auxiliary functions:
\begin{itemize}
\item {\small \FuncSty{consistent($c \in \mathcal{C}$)} $\in
    \{\bot,\top\}$} -- to determine whether the configuration $c$ is
  consistent (cf. Def.~\ref{def:consistent});
\item {\small \FuncSty{subst($c_R \in \mathcal{C}_R$,$c_A \in
      \mathcal{C}_A$)} $\in \{\bot,\top\}$} -- to determine whether
  the configuration $c_R$ is substitutable to $c_A$ (cf.
  Def~\ref{def:structuralRefinement}) ;
\item {\small \FuncSty{enabled($c \in \mathcal{C}$, $R \subseteq
      \mathcal{R}$)} $\subseteq \mathcal{R}$} -- to determine the
  subset of reconfigurations in \AL{$R$} which can be enabled
  from $c$;
\item {\small \FuncSty{pick-up($\mathcal{E} \subseteq \mathcal{R}$)}
    $\in \mathcal{R}$} -- to choose an operation among
  reconfigurations in $\mathcal{E}$;
\item {\small \FuncSty{apply($c \in \mathcal{C}$,$ope \in
      \mathcal{R}$)} $\in \mathcal{C}$} -- to compute the target
  configuration when applying $ope$ to $c$.
\end{itemize}

\begin{wrapfigure}[25]{r}{.48\linewidth}
\vspace{-0.5cm}
  \null\hfill 
  \begin{minipage}{\linewidth}
    \begin{algorithm}[H]
\footnotesize
\SetKwFunction{Struct}{subst}
\SetKwFunction{Enabled}{enabled}
\SetKwFunction{Pick}{pick-up}
\SetKwFunction{Apply}{apply}
\KwData{$c_R^0 \in
\mathcal{C}_R^0$, $c_A^0 \in
\mathcal{C}_A^0$, $\mathcal{R}_R$ and $\mathcal{R}_A$} 
\KwResult{$res \in \{\bot,\top^p\} $, if terminates}
$c_R$ $\leftarrow$ $c_R^0$ \;
$c_A$ $\leftarrow$ $c_A^0$ \;
\While{$\top$}{
  \uIf{\Struct{$c_R$, $c_A$}}{
     $\mathcal{E}_R$ $\leftarrow$ \Enabled{$c_R$, $\mathcal{R}_R$} \;
     $\mathcal{E}_A$ $\leftarrow$ \Enabled{$c_A$, $\mathcal{R}_A$} \;
    \eIf{$\mathcal{E}_R$ $=$ $\emptyset$}{
      \lIf{$\mathcal{E}_A$ $=$ $\emptyset$}{\Return{$res \leftarrow \top^p$}  ;
        \KwSty{break} \nllabel{res:true}  } \;
      \lElse{\Return{$res \leftarrow \bot$} ; \nllabel{res:false:oldpossible} 
         \KwSty{break} \;
         \KwSty{end if}}
    }{
      $ope$ $\leftarrow$ \Pick{$\mathcal{E}_R$} \;
      $c_R$ $\leftarrow$ \Apply{$ope$, $c_R$} \; 
      \lIf{$ope$ $\in$ $\mathcal{R}_R$ $\setminus$ $\mathcal{R}_A$}{ print($\bot^p$) \; \nllabel{res:potential:false} } 
      \Else{
        \uIf{$ope$ $\in$ $\mathcal{R}_R$ $\cap$ $\mathcal{R}_A$ {\bf  and}  $ope$ $\in$ $\mathcal{E}_A$}{
         $c_A$ $\leftarrow$ \Apply{$ope$ ,
          $c_A$ \nllabel{res:potential:true} } \;  print($\top^p$) ;} 
     \lElse{\Return{$res \leftarrow \bot$} ; \nllabel{res:false:opeoldimpossible}
        \KwSty{break} \;
        \KwSty{end if}}
    }
  }
}
   \nllabel{res:false:structural}\lElse{\Return{$res \leftarrow \bot$} ;  \KwSty{break}\;
  \KwSty{end if}}
}%
    \end{algorithm}%
  \end{minipage}%
\hfill \null
\vspace{-0.3cm} \caption{Semi-algorithm on the substitutability\label{fig:algo}}%
\end{wrapfigure}
Let us have a close look at the semi-algorithm. 
\AL{It returns $\bot$ in the following three cases:}
\begin{itemize}
\item Either Line \ref{res:false:structural} indicates that
  clause~\eqref{refinement:0} of Def.~\ref{def:relationraff}
  concerning the structural substitutability from
  Def.~\ref{def:structuralRefinement} is broken.
\item Or Line~\ref{res:false:oldpossible} indicates that there is a
  deadlock at the level after substitutions but not at the level
  before components substitutions. In this case
  clause~\eqref{refinement:4}---the non-introduction of deadlocks---of
  Def.~\ref{def:relationraff} is broken.
\item Or Line ~\ref{res:false:opeoldimpossible} indicates that
  clause~\eqref{refinement:1}---the strict simulation---of
  Def.~\ref{def:relationraff} is broken.
\end{itemize}

%
%
The substitution verification goes on, possibly over infinite
paths. Nevertheless, even in this inconclusive case, the
semi-algorithm can provide some indications on the current status of
the substitutability. Let us consider the set $\mathbb{B}_4 = \{\bot,
\bot^p,\top^p,\top\} $ where $\bot, \top$ stand resp. for {\em{false}}
and {\em{true}} values where as $\bot^p,\top^p$ stand resp. for
{\em{potential false}} and {\em{potential true}} values.  Like for
evaluating temporal properties at runtime as in~\cite{dkl11:ip},
\textit{potential true} and \textit{potential false} values are chosen
whenever an observed behaviour has not yet lead to a violation of the
substitutability-based simulation. With this in mind, when a new
reconfiguration is applied, $\bot^p$ in Line~\ref
{res:potential:false} indicates
\begin{itemize}
\item either a potential trouble with the stuttering simulation:
  clause~\eqref{refinement:2} of Def.~\ref{def:relationraff} may be
  broken if, on the next iteration of the semi-algorithm, the
  structural substitutability---clause~\eqref{refinement:0}---does not
  hold between the configuration reached on the path with
  substitutions and the old configuration on the path before component
  substitutions;
\item or a potential divergence: clause~\eqref{refinement:3} of
  Def.~\ref{def:relationraff} may be broken if no old reconfiguration
  occurs in the future.
\end{itemize}
\noindent When the semi-algorithm indicates $\top^p$, at
Line~\ref{res:potential:true}, it means that the clauses of
Def.~\ref{def:relationraff} have not yet been violated, and the
verification of the substitutability-based simulation must continue.

\AL{Finally, when the semi-algorithm terminates and returns $\top^p$
  (line~\ref{res:true}), it indicates that \emph{finite} paths
  have been considered and 
  no more reconfigurations can be fired at both pre- and
  post-substitution levels. \OK{It means that until this point all clauses of
  Def.~\ref{def:relationraff} are satisfied. This information can be exploited for semi-deciding 
  the substitutability on other reconfigurations sequences.}}

\begin{proposition}
\label{prop:decision}
Given $S_A$ and $S_R$, if the substitutability semi-algorithm
terminates by providing the $\bot$ value then one has $S_R
\not\sqsubseteq_{\FuncSty{subst}} S_A$.
\end{proposition} 
The idea behind Proposition~\ref{prop:decision} is as follows: if
there are two sequences of dynamic reconfigurations on which one of
the substitutability relation clauses is violated then it does imply
the substitutability-based simulation violation.

\medskip \AL{Figure~\ref{fig:example-subs} illustrates the application of the substitutability semi-algorithm. 
When a new reconfiguration is
executed (leading for example to $14$ linked to $02$), the evaluation
gives $\bot^p$, although the structural substitutability holds. It is
due to the fact that the new reconfigurations may take control
forever, depending of course on future reconfigurations. In contrast,
when an old reconfiguration is executed (leading for example to $15$
which is linked to $03$), the evaluation becomes $\top^p$: the
structural substitutability holds and the potential livelock has been
avoided.  Consequently, when considering finite parts of paths in
Fig.~\ref{fig:example-subs} until the current pair $(c_R, c_A)$, the
reconfigurations of the HTTP server combine well with reconfigurations
due to component substitutions.}


\section{Experiments}
\label{sec:experimentation}
This section provides a proof of concept by reporting on
experiments using the B tools to express and to check the
consistency and substitutability constraints, and to implement the
substitutability semi-algorithm.

\begin{figure}[ht]
  \centering
  \includegraphics[width=.9\linewidth]{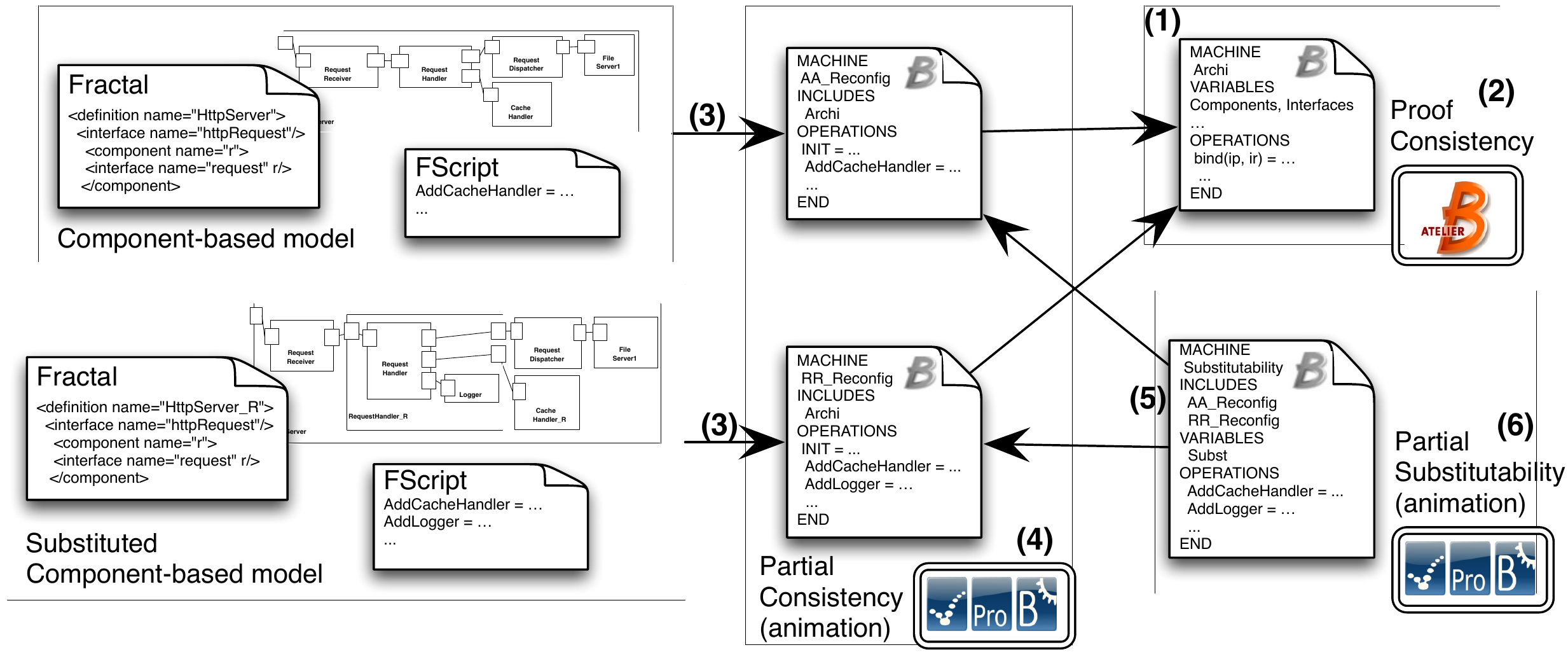}
  \caption{Principle of the validation framework}
  \label{principle}
\end{figure}

\subsection{A Formal Toolset: the B Method}

B is a formal software development method used to model systems and to
reason about their development~\cite{BBook}.  When building a B
machine, the principle is to express system
properties---invariants---which are always true after each evolution
step of the machine, the evolution being specified by the B
operations. The verification of a machine correctness is thus akin to
verifying the preservation of these properties, no matter which step
of evolution the system takes.

The B method is based on set theory, relations and first-order logic.
Constraints are specified in the \B{INVARIANT} clause of the machine,
and its evolution is specified by operations in the \B{OPERATIONS}
clause. Let us assume here that the initialisation is a special kind
of operation. In this setting, the {\em consistency} checking of a B
machine consists in verifying that each operation satisfies the
\B{INVARIANT} assuming its precondition and the invariant hold.


The tools, such as B4free or
AtelierB\footnote{Available at \url{http://www.b4free.com} or
  \url{http://www.atelierb.eu}, }, automatically generate proof
obligations (POs) to ensure the consistency in the sense of
B~\cite{BBook}. Some of them are obvious POs whereas the other POs
have to be proved interactively if it was not done fully automatically
by the different provers embedded into AtelierB.
Another tool, called
ProB\footnote{Available at \url{http://www.stups.uni-duesseldorf.de/ProB}}, allows
the user to animate B machines for their debugging and testing. On the
verification side, ProB contains a constraint-based checker and a LTL
bounded model-checker with particular features; Both checkers can be used to
validate B machines~\cite{LeuschelB03,LeuschelP07}.

\subsection{Consistency Checking by Proof and Model Animation}

This section summarises the work in~\cite{lanoix11a} on
specifications in B of the proposed component-based model with
reconfigurations, and on verification process using the B tools, by
combining proof and model-checking techniques.  Let us consider the B
machines which, for readability reasons, are simplified versions of
the "real" B machines.

The configuration model given in Def.~\ref{def:config}
(appendix~\ref{appendix:config}) can be easily translated into a B
machine \B{Archi} ({\bf(1)} in Fig.~\ref{principle}).  In this
machine, the sets as \B{Components} or \B{Interfaces}, and relations
as \B{Parent} or \B{Binding} are defined into the \B{VARIABLES}
clause; the architectural consistency constraints $CC$ are defined
into the \B{INVARIANT} clause; the basic reconfigurations operations
as $bind(ip,ir)$ or $start(compo)$ are also defined here as B
operations. Then, we use the AtelierB tool to interactively
demonstrate the consistency of the architectural constraints ({\bf(2)}
in Fig.~\ref{principle}) through the basic reconfiguration operations.

\begin{lstlisting}[frame=single]
MACHINE
 Archi
VARIABLES
 Components, Interfaces, ProvInterfaces, ReqInterfaces, Supplier, Parent, Binding, ...
INVARIANT
   ProvInterfaces <: Interfaces & ReqInterfaces <: Interfaces
 & ProvInterfaces \/ ReqInterfaces = Interfaces & ProvInterfaces /\ ReqInterfaces = {}
 & Supplier : Interfaces --> Components
 & Parent : Components <-> Components
 & Binding : ProvInterfaces +-> ReqInterfaces
 & closure1(Parent) /\ id(Components) = {}			                          			    /* CC.3 */        
 & ! (ip,ir).(ip |-> ir : Binding => Provider(ip) /= Requirer(ir) & Parent(Supplier(iprov)) = Parent(Supplier(ireq)) ) /* CC.4 + CC.5 */ 
 & ...
OPERATIONS
bind(ip,ir) = 
PRE
    ip : ProvInterfaces & ir : ReqInterfaces & ip|->ir /: Binding & ip /: dom(Binding) & ip /: dom(Delegate) & ir /: dom(Delegate)
THEN 
    Binding(ip) := ir 
END ; 
 ...
END
\end{lstlisting}

Then, the generic B machine \B{Archi} is instantiated as \B{Reconfig}
to represent an architecture under consideration, particularly by
giving values to all the sets and relations to represent the
considered component architecture configuration and by implementing
the non-primitive reconfiguration operations using the basic ones
({\bf(3)} in Fig.~\ref{principle}).  At this point, we can perform a
(partial) validation of the instantiated B machine \B{Reconfig}
through animations, thanks to the ProB model-checker features
({\bf(4)} in Fig.~\ref{principle}).

\begin{lstlisting}[frame=single]
MACHINE
 Reconfig
INCLUDES
 Archi
OPERATIONS
INIT =
 BEGIN
  Components := { HttpServer, RequestReceiver, RequestHandler, CacheHandler, RequestDispatcher, FileServer1, FileServer2 } 
  || ProvInterfaces := { httpRequest, request, handler, cache, dispatcher, server1, server2 }
  || ReqInterfaces := { getHandler, getDispatcher, getCache, getServer }
  || Parent := { RequestReceiver|->HttpServer, RequestHandler|->HttpServer, CacheHandler|->HttpServer, RequestDispatcher|->HttpServer }
  || Binding := { handler|->getHandler, cache|->getCache, dispatcher|->getDispatcher, server1|->getServer }
  ...
 END ;
 AddCacheHandler = 
 BEGIN 
  instantiate(CacheHandler) ;
  add(CacheHandler, HttpServer) ;
  bind(cache, getCache) ;
  start(CacheHandler)
 END ;
...
END
\end{lstlisting}

\subsection{Substitutability Checking by Model Animation}

We exploit the work in~\cite{lanoix11a} by considering two
instantiated B models \B{AA_Reconfig} and \B{RR_Reconfig} which define
two component architectures, wrt. the pre-/post-substitution
levels. All the elements and relations are defined twice:
\B{AA.Components}, \B{RR.Components}, \B{AA.Interfaces},
\B{RR.Interfaces}, \B{AA.Parent} or \B{RR.Parent} \dots A new machine
\B{Substitutability} includes these two models ({\bf(5)} in
Fig.~\ref{principle}).  It defines the \emph{substitute
  reconfiguration function} \B{Subst} to link together the
\B{AA.Components} to the substituted \B{RR.Components}.

\begin{lstlisting}[frame=single]
MACHINE
 Substitutability
INCLUDES
 AA_Reconfig
 RR_Reconfig
VARIABLES
 Subst
INVARIANT
 Subst : AA.Components +-> RR.Components
 & !(c,i).(cc : AA.Components /\ RR.Components & i : AA.Interfaces & AA.Supplier(i) = cc => RR.Supplier(i) = c)     /* SC.5 */ 
 & !(ca).(AA.Components - RR.Components  => #(cr).(RR.Components - AA.Components & Subst(ca) = cr))              /* SC.7 */
 & AA.Interfaces <: RR.Interfaces & AA.ProvInterfaces <: RR.ProvInterfaces & AA.ReqInterfaces <: RR.ReqInterfaces     /* SC.13 */
 & !(i).(i : RR.ProvInterfaces - AA.ProvInterfaces => RR.Supplier(i) : RR.Components - AA.Components)             /* SC.17 */
 ...
INITIALISATION
 Subst := {CacheHandler|->CacheHandlerR, RequestHandler|->RequestHandlerR}
OPERATIONS
 AddCacheHandler = 
 BEGIN
  AA_AddCacheHandler || RR_AddCacheHandler
 END;
 AddLogger = 
 BEGIN
  RR_AddLogger
 END ;
 ...
END
\end{lstlisting}

The architectural substitutability constraints $SC_{Subst}$ are
defined into the \B{INVARIANT} clause of this machine; they are
constraints between the elements and relations of \B{AA_Reconfig}, and
the elements and relations of \B{RR_Reconfig}. For example, the reader
can see some clauses expressed above as a part of the \B{INVARIANT}.

 \begin{figure}[ht]
   \centering
   \includegraphics[width=.99\linewidth]{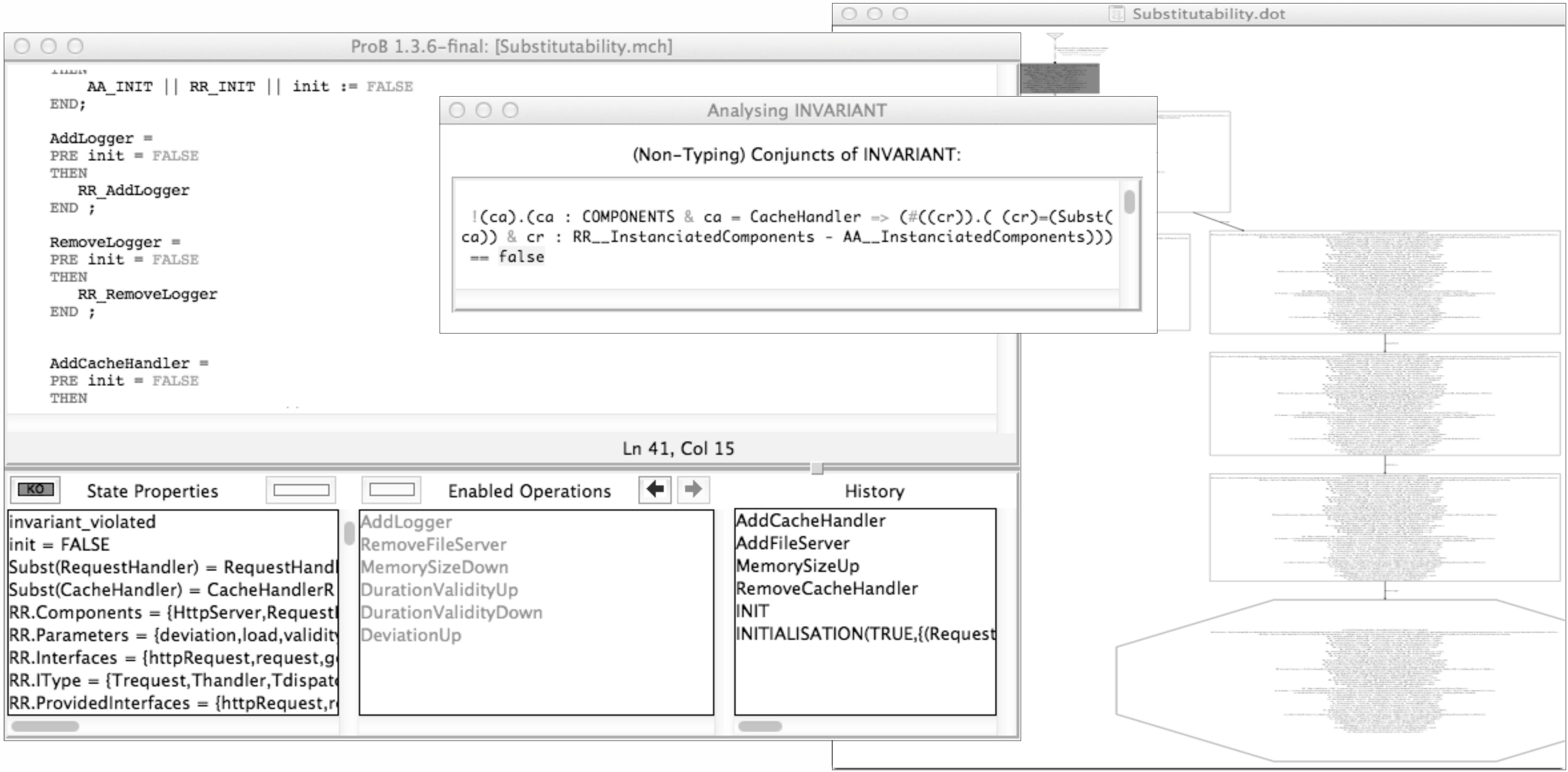}
   \caption{The ProB tool: invariant broken illustrating
    substitutability constraints broken}
   \label{fig:prob}
 \end{figure}

Afterwards, we use the ProB model-checker to animate the
\B{Substitutability} machine and to explore---simultaneously--- the
two instantiated B models, i.e. the pre-/post-substitution component architectures
({\bf(6)} in Fig.~\ref{principle}). This animation allows us to perform the
evaluations needed for the semi-algorithm from
Section~\ref{sec:relSubs}: we choose the next
dynamic reconfiguration to be applied on the ``Enabled operations'' windows
of ProB (see Fig. ~\ref{fig:prob}); if it is an old reconfiguration
operation, it is simultaneously executed into \B{AA_Reconfig} and
\B{RR_Reconfig}, otherwise it is only run into \B{RR_Reconfig}; then,
the \B{INVARIANT} checking corresponds to the validation of all the
$SC_{Subst}$ constraints.

 Let us suppose that after the reconfiguration by component
 substitution the \reconf{AddCacheHandler} dynamic reconfiguration
 contains an implementation error: it does not add the
 \compo{CacheHandler\_R} component.  When using ProB, we have easily
 found the error. Indeed, when \reconf{AddCacheHandler} is executed
 simultaneously by \B{AA_Reconfig} and \B{RR_Reconfig}, the invariant
 is broken as depicted on Fig.~\ref{fig:prob}. More precisely, the
 correspondinf clause into $SC_{Subst}$ is broken, as the
 \compo{CacheHandler} component has no substituted component
 w.r.t. the $Subst$ function.


\section{Discussion and Conclusion}
\label{sec:conclu}
{\em Related work.}  For distributed components like Fractal, GCM and
ProActive components, the role of automata-based analysis providing a
formal basis for automatic tool support is emphasised
in~\cite{DBLP:journals/adt/BarrosACHM09}.  In the context of dynamic
reconfigurations, ArchJava~\cite{Aldric08} gives means to reconfigure
Java architectures, and to guarantee communication integrity at
run-time. In~\cite{BarringerGR07} a temporal logic based framework to
deal with systems evolution is proposed.

To compare processes or components, the bisimulation equivalence by
Milner~\cite{milner80} and Park~\cite{park81} is widely used: It
preserves branching behaviours and, consequently, most of the dynamic
properties; there is a link between the strong bisimulation and modal
logics~\cite{hennessy85}; this is a congruence for a number of
composition operators. There are numerous works dealing with component
substitutability or
interoperability~\cite{DBLP:conf/cbse/2007,DBLP:journals/entcs/CernaVZ07,DBLP:conf/euromicro/BradaV06}.
Our work is close to that in~\cite{DBLP:journals/entcs/CernaVZ07},
where a component substitutability is defined using equivalences
between component-interaction automata wrt. a given set of observable
labels.  In the present work, in addition to a set of labels,
divergency, livelocks are taken into account when comparing execution
paths.  As KLAPER~\cite{DBLP:conf/dagstuhl/GrassiMRS07},
Palladio~\cite{DBLP:conf/wosp/BeckerKR07} and RoboCop
\cite{DBLP:conf/euromicro/FioukovEHC02} component models do not define
any refinement/substitution notion, they are clearly distinguishable
from our work.

Let us remark that the substitutability-based simulation in this paper
is close to the refinement relation in~\cite{dormoy12a}.  However, as
\cite{dormoy12a} focuses on a linear temporal logic property
preservation, no method is given in~\cite{dormoy12a} to verify the
structural refinement.

{\em Conclusion.}  This paper extends the previous work on the
consistency verification of the component-based architectures by
introducing a new reconfiguration operation based on components
substitutions, and by integrating it into a simulation relation
handling dynamic reconfigurations.  A semi-algorithm is proposed to
evaluate on the fly the substitutability relation and its partial
correctness is established.  As a proof of concept, the B tools are
used for dealing with the substitutability constraints through dynamic
reconfigurations.  As the ProB tool can deal with a dialect of linear
temporal logic, we intend to accompany the present work on component
substitutability with a runtime (bounded) model-checking of linear
temporal logic patterns.  Further, we plan to combine our results with
adaptation policies: the partial evaluations $\bot^p$ and $\top^p$
could \OK{be taken into account within the adaption policies
framework, to choose the most appropriate reconfiguration to be applied to the system under scrutiny.}


\bibliographystyle{eptcs}
\bibliography{main}

\begin{thebibliography}{10}
\providecommand{\bibitemdeclare}[2]{}
\providecommand{\surnamestart}{}
\providecommand{\surnameend}{}
\providecommand{\urlprefix}{Available at }
\providecommand{\url}[1]{\texttt{#1}}
\providecommand{\href}[2]{\texttt{#2}}
\providecommand{\urlalt}[2]{\href{#1}{#2}}
\providecommand{\doi}[1]{doi:\urlalt{http://dx.doi.org/#1}{#1}}
\providecommand{\bibinfo}[2]{#2}

\bibitemdeclare{book}{BBook}
\bibitem{BBook}
\bibinfo{author}{J.-R. \surnamestart Abrial\surnameend} (\bibinfo{year}{1996}):
  \emph{\bibinfo{title}{The {B}~Book - Assigning Programs to Meanings}}.
\newblock \bibinfo{publisher}{Cambridge University Press},
  \doi{10.1017/CBO9780511624162}.

\bibitemdeclare{article}{nazareno02a}
\bibitem{nazareno02a}
\bibinfo{author}{N.~\surnamestart Aguirre\surnameend} \&
  \bibinfo{author}{T.~\surnamestart Maibaum\surnameend} (\bibinfo{year}{2002}):
  \emph{\bibinfo{title}{A Temporal Logic Approach to the Specification of
  Reconfigurable Component-Based Systems}}.
\newblock {\sl \bibinfo{journal}{Automated Software Engineering}},
  \doi{10.1109/ASE.2002.1115028}.

\bibitemdeclare{inproceedings}{Aldric08}
\bibitem{Aldric08}
\bibinfo{author}{J.~\surnamestart Aldric\surnameend} (\bibinfo{year}{2008}):
  \emph{\bibinfo{title}{Using Types to Enforce Architectural Structure}}.
\newblock In: {\sl \bibinfo{booktitle}{WICSA'08}}, pp. \bibinfo{pages}{23--34},
  \doi{10.1109/WICSA.2008.48}.

\bibitemdeclare{inproceedings}{BarringerGR07}
\bibitem{BarringerGR07}
\bibinfo{author}{H.~\surnamestart Barringer\surnameend}, \bibinfo{author}{D.~M.
  \surnamestart Gabbay\surnameend} \& \bibinfo{author}{D.~E. \surnamestart
  Rydeheard\surnameend} (\bibinfo{year}{2007}): \emph{\bibinfo{title}{From
  Runtime Verification to Evolvable Systems}}.
\newblock In: {\sl \bibinfo{booktitle}{RV}}, {\sl \bibinfo{series}{LNCS}}
  \bibinfo{volume}{4839}, \bibinfo{publisher}{Springer}, pp.
  \bibinfo{pages}{97--110}, \doi{10.1007/978-3-540-77395-5\_9}.

\bibitemdeclare{article}{DBLP:journals/adt/BarrosACHM09}
\bibitem{DBLP:journals/adt/BarrosACHM09}
\bibinfo{author}{T.~\surnamestart Barros\surnameend},
  \bibinfo{author}{R.~\surnamestart Ameur-Boulifa\surnameend},
  \bibinfo{author}{A.~\surnamestart Cansado\surnameend},
  \bibinfo{author}{L.~\surnamestart Henrio\surnameend} \&
  \bibinfo{author}{E.~\surnamestart Madelaine\surnameend}
  (\bibinfo{year}{2009}): \emph{\bibinfo{title}{Behavioural models for
  distributed {Fractal} components}}.
\newblock {\sl \bibinfo{journal}{Annales des T{\'e}l{\'e}communications}}
  \bibinfo{volume}{64}(\bibinfo{number}{1-2}), pp. \bibinfo{pages}{25--43},
  \doi{10.1007/s12243-008-0069-7}.

\bibitemdeclare{inproceedings}{DBLP:conf/wosp/BeckerKR07}
\bibitem{DBLP:conf/wosp/BeckerKR07}
\bibinfo{author}{S.~\surnamestart Becker\surnameend},
  \bibinfo{author}{H.~\surnamestart Koziolek\surnameend} \&
  \bibinfo{author}{R.~\surnamestart Reussner\surnameend}
  (\bibinfo{year}{2007}): \emph{\bibinfo{title}{Model-Based performance
  prediction with the palladio component model}}.
\newblock In: {\sl \bibinfo{booktitle}{Proceedings of the 6th International
  Workshop on Software and Performance, WOSP 2007}}, \bibinfo{publisher}{ACM},
  pp. \bibinfo{pages}{54--65}, \doi{10.1145/1216993.1217006}.

\bibitemdeclare{inproceedings}{DBLP:conf/euromicro/BradaV06}
\bibitem{DBLP:conf/euromicro/BradaV06}
\bibinfo{author}{P.~\surnamestart Brada\surnameend} \&
  \bibinfo{author}{L.~\surnamestart Valenta\surnameend} (\bibinfo{year}{2006}):
  \emph{\bibinfo{title}{Practical Verification of Component Substitutability
  Using Subtype Relation}}.
\newblock In: {\sl \bibinfo{booktitle}{32nd EUROMICRO Conference on Software
  Engineering and Advanced Applications, EUROMICRO-SEAA 2006}},
  \bibinfo{publisher}{IEEE}, pp. \bibinfo{pages}{38--45},
  \doi{10.1109/EUROMICRO.2006.50}.

\bibitemdeclare{article}{Butler96}
\bibitem{Butler96}
\bibinfo{author}{M.~J. \surnamestart Butler\surnameend} (\bibinfo{year}{1996}):
  \emph{\bibinfo{title}{Stepwise Refinement of Communicating Systems}}.
\newblock {\sl \bibinfo{journal}{Sci. Comput. Program.}}
  \bibinfo{volume}{27}(\bibinfo{number}{2}), pp. \bibinfo{pages}{139--173},
  \doi{10.1016/0167-6423(96)81173-7}.

\bibitemdeclare{article}{DBLP:journals/entcs/CernaVZ07}
\bibitem{DBLP:journals/entcs/CernaVZ07}
\bibinfo{author}{I.~\surnamestart Cern{\'a}\surnameend},
  \bibinfo{author}{P.~\surnamestart Varekov{\'a}\surnameend} \&
  \bibinfo{author}{B.~\surnamestart Zimmerova\surnameend}
  (\bibinfo{year}{2007}): \emph{\bibinfo{title}{Component Substitutability via
  Equivalencies of Component-Interaction Automata}}.
\newblock {\sl \bibinfo{journal}{Electr. Notes Theor. Comput. Sci.}}
  \bibinfo{volume}{182}, pp. \bibinfo{pages}{39--55},
  \doi{10.1016/j.entcs.2006.09.030}.

\bibitemdeclare{article}{colin07a}
\bibitem{colin07a}
\bibinfo{author}{S.~\surnamestart Colin\surnameend},
  \bibinfo{author}{A.~\surnamestart Lanoix\surnameend} \&
  \bibinfo{author}{J.~\surnamestart Souqui{\`e}res\surnameend}
  (\bibinfo{year}{2009}): \emph{\bibinfo{title}{Trustworthy interface
  compliancy: data model adaptation}}.
\newblock {\sl \bibinfo{journal}{Electronic Notes in Theoretical Computer
  Science}} \bibinfo{volume}{203}(\bibinfo{number}{7}), pp.
  \bibinfo{pages}{23--35}, \doi{10.1016/j.entcs.2009.03.024}.

\bibitemdeclare{inproceedings}{aguilar01a}
\bibitem{aguilar01a}
\bibinfo{author}{M.~Aguilar \surnamestart Cornejo\surnameend},
  \bibinfo{author}{H.~\surnamestart Garavel\surnameend},
  \bibinfo{author}{R.~\surnamestart Mateescu\surnameend} \&
  \bibinfo{author}{N.~De \surnamestart Palma\surnameend}
  (\bibinfo{year}{2001}): \emph{\bibinfo{title}{Specification and Verification
  of a Dynamic Reconfiguration Protocol for Agent-Based Applications}}.
\newblock In: {\sl \bibinfo{booktitle}{DAIS}}, pp. \bibinfo{pages}{229--244}.

\bibitemdeclare{inproceedings}{dormoy10b}
\bibitem{dormoy10b}
\bibinfo{author}{J.~\surnamestart Dormoy\surnameend},
  \bibinfo{author}{O.~\surnamestart Kouchnarenko\surnameend} \&
  \bibinfo{author}{A.~\surnamestart Lanoix\surnameend} (\bibinfo{year}{2010}):
  \emph{\bibinfo{title}{Using Temporal Logic for Dynamic Reconfigurations of
  Components}}.
\newblock In: {\sl \bibinfo{booktitle}{FACS 2010, 7th Int. Ws. on Formal
  Aspects of Component Software}}, {\sl \bibinfo{series}{LNCS}}
  \bibinfo{volume}{6921}, \bibinfo{publisher}{Springer}, pp.
  \bibinfo{pages}{200--217}, \doi{10.1007/978-3-642-27269-1\_12}.

\bibitemdeclare{inproceedings}{dkl11:ip}
\bibitem{dkl11:ip}
\bibinfo{author}{J.~\surnamestart Dormoy\surnameend},
  \bibinfo{author}{O.~\surnamestart Kouchnarenko\surnameend} \&
  \bibinfo{author}{A.~\surnamestart Lanoix\surnameend} (\bibinfo{year}{2011}):
  \emph{\bibinfo{title}{Runtime Verification of Temporal Patterns for Dynamic
  Reconfigurations of Components}}.
\newblock In: {\sl \bibinfo{booktitle}{FACS 2011}}, {\sl
  \bibinfo{series}{LNCS}} \bibinfo{volume}{7253},
  \bibinfo{publisher}{Springer}, pp. \bibinfo{pages}{115--132},
  \doi{10.1007/978-3-642-35743-5\_8}.

\bibitemdeclare{inproceedings}{dormoy12a}
\bibitem{dormoy12a}
\bibinfo{author}{J.~\surnamestart Dormoy\surnameend},
  \bibinfo{author}{O.~\surnamestart Kouchnarenko\surnameend} \&
  \bibinfo{author}{A.~\surnamestart Lanoix\surnameend} (\bibinfo{year}{2012}):
  \emph{\bibinfo{title}{{When Structural Refinement of Components Keeps
  Temporal Properties Over Reconfigurations}}}.
\newblock In: {\sl \bibinfo{booktitle}{18th International Symposium on Formal
  Methods (FM 2012)}}, {\sl \bibinfo{series}{LNCS}} \bibinfo{volume}{7436},
  \bibinfo{publisher}{Springer-Verlag}, \doi{10.1007/978-3-642-32759-9\_16}.

\bibitemdeclare{inproceedings}{DBLP:conf/euromicro/FioukovEHC02}
\bibitem{DBLP:conf/euromicro/FioukovEHC02}
\bibinfo{author}{A.~V. \surnamestart Fioukov\surnameend}, \bibinfo{author}{E.M.
  \surnamestart Eskenazi\surnameend}, \bibinfo{author}{D.~K. \surnamestart
  Hammer\surnameend} \& \bibinfo{author}{M.~R.~V. \surnamestart
  Chaudron\surnameend} (\bibinfo{year}{2002}): \emph{\bibinfo{title}{Evaluation
  of Static Properties for Component-Based Architectures}}.
\newblock In: {\sl \bibinfo{booktitle}{28th EUROMICRO Conference 2002}},
  \bibinfo{publisher}{IEEE Computer Society}, pp. \bibinfo{pages}{33--39}.
\newblock
  \urlprefix\url{http://computer.org/proceedings/euromicro/1787/17870033abs.ht%
m}.

\bibitemdeclare{inproceedings}{Glabbeek93}
\bibitem{Glabbeek93}
\bibinfo{author}{R.~J. \surnamestart van Glabbeek\surnameend}
  (\bibinfo{year}{1993}): \emph{\bibinfo{title}{The Linear Time - Branching
  Time Spectrum II}}.
\newblock In: {\sl \bibinfo{booktitle}{CONCUR '93, 4th International Conference
  on Concurrency Theory}}, {\sl \bibinfo{series}{LNCS}} \bibinfo{volume}{715},
  \bibinfo{publisher}{Springer}, pp. \bibinfo{pages}{66--81},
  \doi{10.1007/3-540-57208-2\_6}.

\bibitemdeclare{inproceedings}{DBLP:conf/dagstuhl/GrassiMRS07}
\bibitem{DBLP:conf/dagstuhl/GrassiMRS07}
\bibinfo{author}{V.~\surnamestart Grassi\surnameend},
  \bibinfo{author}{R.~\surnamestart Mirandola\surnameend},
  \bibinfo{author}{E.~\surnamestart Randazzo\surnameend} \&
  \bibinfo{author}{A.~\surnamestart Sabetta\surnameend} (\bibinfo{year}{2007}):
  \emph{\bibinfo{title}{KLAPER: An Intermediate Language for Model-Driven
  Predictive Analysis of Performance and Reliability}}.
\newblock In: {\sl \bibinfo{booktitle}{The Common Component Modeling Example:
  Comparing Software Component Models}}, {\sl \bibinfo{series}{LNCS}}
  \bibinfo{volume}{5153}, \bibinfo{publisher}{Springer}, pp.
  \bibinfo{pages}{327--356}, \doi{10.1007/978-3-540-85289-6\_13}.

\bibitemdeclare{book}{Hamilton78}
\bibitem{Hamilton78}
\bibinfo{author}{A.~G. \surnamestart Hamilton\surnameend}
  (\bibinfo{year}{1978}): \emph{\bibinfo{title}{Logic for mathematicians}}.
\newblock \bibinfo{publisher}{Cambridge University Press, Cambridge}.

\bibitemdeclare{article}{hennessy85}
\bibitem{hennessy85}
\bibinfo{author}{M.~\surnamestart Hennessy\surnameend} \&
  \bibinfo{author}{R.~\surnamestart Milner\surnameend} (\bibinfo{year}{1985}):
  \emph{\bibinfo{title}{Algebraic Laws for Nondeterminism and Concurrency}}.
\newblock {\sl \bibinfo{journal}{Journal of the ACM}}
  \bibinfo{volume}{32}(\bibinfo{number}{1}), pp. \bibinfo{pages}{137--161},
  \doi{10.1145/2455.2460}.

\bibitemdeclare{inproceedings}{KestenMP93}
\bibitem{KestenMP93}
\bibinfo{author}{Y.~\surnamestart Kesten\surnameend},
  \bibinfo{author}{Z.~\surnamestart Manna\surnameend} \&
  \bibinfo{author}{A.~\surnamestart Pnueli\surnameend} (\bibinfo{year}{1994}):
  \emph{\bibinfo{title}{Temporal Verification of Simulation and Refinement}}.
\newblock In: {\sl \bibinfo{booktitle}{A Decade of Concurrency, Reflections and
  Perspectives, REX School/Symposium}}, {\sl \bibinfo{series}{LNCS}}
  \bibinfo{volume}{803}, \bibinfo{publisher}{Springer}, pp.
  \bibinfo{pages}{273--346}, \doi{10.1007/3-540-58043-3\_22}.

\bibitemdeclare{inproceedings}{lanoix11a}
\bibitem{lanoix11a}
\bibinfo{author}{A.~\surnamestart Lanoix\surnameend},
  \bibinfo{author}{J.~\surnamestart Dormoy\surnameend} \&
  \bibinfo{author}{O.~\surnamestart Kouchnarenko\surnameend}
  (\bibinfo{year}{2011}): \emph{\bibinfo{title}{Combining Proof and
  Model-checking to Validate Reconfigurable Architectures}}.
\newblock In: {\sl \bibinfo{booktitle}{FESCA 2011}}, \bibinfo{series}{ENTCS},
  \doi{10.1016/j.entcs.2011.11.011}.

\bibitemdeclare{article}{lanoix07b}
\bibitem{lanoix07b}
\bibinfo{author}{A.~\surnamestart Lanoix\surnameend} \&
  \bibinfo{author}{J.~\surnamestart Souqui{\`e}res\surnameend}
  (\bibinfo{year}{2008}): \emph{\bibinfo{title}{A Trustworthy Assembly of
  Components using the {B} Refinement}}.
\newblock {\sl \bibinfo{journal}{e-Informatica Software Engineering Journal
  (ISEJ)}} \bibinfo{volume}{2}(\bibinfo{number}{1}), pp.
  \bibinfo{pages}{9--28}.
\newblock
  \urlprefix\url{http://www.e-informatyka.pl/attach/e-Informatica\_-\_Volume_2%
/Vol2Iss1Art1eInformatica.pdf}.

\bibitemdeclare{inproceedings}{leger10a}
\bibitem{leger10a}
\bibinfo{author}{M.~\surnamestart L{\'e}ger\surnameend}, \bibinfo{author}{Th.
  \surnamestart Ledoux\surnameend} \& \bibinfo{author}{Th. \surnamestart
  Coupaye\surnameend} (\bibinfo{year}{2010}): \emph{\bibinfo{title}{Reliable
  Dynamic Reconfigurations in a Reflective Component Model}}.
\newblock In: {\sl \bibinfo{booktitle}{CBSE 2010}}, {\sl
  \bibinfo{series}{LNCS}} \bibinfo{volume}{6092}, pp. \bibinfo{pages}{74--92},
  \doi{10.1007/978-3-642-13238-4\_5}.

\bibitemdeclare{inproceedings}{LeuschelB03}
\bibitem{LeuschelB03}
\bibinfo{author}{M.~\surnamestart Leuschel\surnameend} \&
  \bibinfo{author}{M.~J. \surnamestart Butler\surnameend}
  (\bibinfo{year}{2003}): \emph{\bibinfo{title}{{ProB}: A Model Checker for
  {B}}}.
\newblock In: {\sl \bibinfo{booktitle}{Int. Symp. of Formal Methods Europe
  FME'03}}, {\sl \bibinfo{series}{LNCS}} \bibinfo{volume}{2805},
  \bibinfo{publisher}{Springer}, pp. \bibinfo{pages}{855--874},
  \doi{10.1007/978-3-540-45236-2\_46}.

\bibitemdeclare{inproceedings}{LeuschelP07}
\bibitem{LeuschelP07}
\bibinfo{author}{M.~\surnamestart Leuschel\surnameend} \&
  \bibinfo{author}{D.~\surnamestart Plagge\surnameend} (\bibinfo{year}{2007}):
  \emph{\bibinfo{title}{Seven at one stroke: LTL model checking for High-level
  Specifications in B, Z, CSP, and more}}.
\newblock In: {\sl \bibinfo{booktitle}{ISoLA'07}}, {\sl \bibinfo{series}{Revue
  des Nouvelles Technologies de l'Information}} \bibinfo{volume}{RNTI-SM-1},
  pp. \bibinfo{pages}{73--84}.

\bibitemdeclare{book}{milner80}
\bibitem{milner80}
\bibinfo{author}{R.~\surnamestart Milner\surnameend} (\bibinfo{year}{1980}):
  \emph{\bibinfo{title}{A Calculus of Communicating Systems}}.
\newblock {\sl \bibinfo{series}{Lecture Notes in Computer
  Science}}~\bibinfo{volume}{92}, \bibinfo{publisher}{Springer Verlag},
  \doi{10.1007/3-540-10235-3}.

\bibitemdeclare{book}{Milner:1989}
\bibitem{Milner:1989}
\bibinfo{author}{R.~\surnamestart Milner\surnameend} (\bibinfo{year}{1989}):
  \emph{\bibinfo{title}{Communication and Concurrency}}.
\newblock \bibinfo{publisher}{Prentice-Hall, Inc.}

\bibitemdeclare{inproceedings}{park81}
\bibitem{park81}
\bibinfo{author}{D.~\surnamestart Park\surnameend} (\bibinfo{year}{1981}):
  \emph{\bibinfo{title}{Concurrency and Automata on Infinite Sequences}}.
\newblock In: {\sl \bibinfo{booktitle}{Lecture Notes in Computer Science}},
  \bibinfo{volume}{104}, \bibinfo{publisher}{Springer Verlag}, pp.
  \bibinfo{pages}{167--183}, \doi{10.1007/BFb0017309}.

\bibitemdeclare{proceedings}{DBLP:conf/cbse/2007}
\bibitem{DBLP:conf/cbse/2007}
\bibinfo{editor}{H.~W. \surnamestart Schmidt\surnameend},
  \bibinfo{editor}{I.~\surnamestart Crnkovic\surnameend},
  \bibinfo{editor}{G.~T. \surnamestart Heineman\surnameend} \&
  \bibinfo{editor}{J.~A. \surnamestart Stafford\surnameend}, editors
  (\bibinfo{year}{2007}): \emph{\bibinfo{title}{Component-Based Software
  Engineering, 10th International Symposium, CBSE 2007, Medford, MA, USA, July
  9-11, 2007, Proceedings}}. {\sl \bibinfo{series}{LNCS}}
  \bibinfo{volume}{4608}, \bibinfo{publisher}{Springer},
  \doi{10.1007/978-3-540-73551-9}.

\end{thebibliography}

\appendix

 \newpage
 \section{Architectural Configuration Definition~\cite{dkl11:ip}}
 \label{appendix:config}
 \begin{definition}[Configuration]\label{def:config}
  A configuration $c$ is a tuple $\langle Elem, Rel \rangle$ where
  \begin{itemize}
  \item $Elem = Components\ \uplus\ Interfaces\ \uplus\ Parameters\
    \uplus\ Types $ is a set of architectural elements,
    such that
    \begin{itemize}
    \item $Components$ is a non-empty set of the core entities, i.e
      components;
    \item $Interfaces = ReqInterfaces \uplus ProvInterfaces$
      is a finite set of the (required and provided) interfaces;
    \item $Parameters$ is a finite set of component parameters;
    \item $Types = ITypes \uplus\ PTypes$ is a finite set of the
interface types and the parameter data types;
    \end{itemize}
  \item 
	$Rel = \left\{\begin{array}{l}
        Container\ \uplus\ ContainerType\ \uplus\  Parent \\
        \uplus\ Binding\ \uplus\  Delegate\ \uplus\ State \ \uplus\   Value
      \end{array} \right. $ 

    is a set of architectural relations which link architectural
    elements, such that
    \begin{itemize}
    \item $Container\, :\, Interfaces\, \uplus\, Parameters\rightarrow
      Components$ is a \emph{total function} giving the component
      which supplies the considered interface or the component of a
      considered parameter;
    \item $ContainerType\, :\, Interfaces\, \uplus\, Parameters
      \rightarrow Types$ is a \emph{total function} that associates a
      type with each required/provided interface, or with a parameter;
    \item $Parent \subseteq Components \times Components$ is a
      \emph{relation} linking a sub-component to the corresponding
      composite component\footnote{For any $(p,q)\in Parent$, we say
        that $q$ has a sub-component $p$, i.e. $p$ is a child of
        $q$.};
    \item $Binding\, :\, ProvInterfaces \rightarrow
      ReqInterfaces$ is a \emph{partial function} which binds
      together a provided interface and a required one;
    \item $Delegate\, :\, Interfaces \rightarrow Interfaces$ is a
      \emph{partial function} which expresses delegation links;
    \item $State\, :\, Components \rightarrow \{started, stopped\}$ is
      a \emph{total function} giving the status of instantiated
      components;
    \item $Contingency\, :\, ReqInterfaces \rightarrow
      \{mandatory,optional\}$ is a \emph{total function} to
      characterise the required interfaces;
    \item $Value\, :\, Parameters \rightarrow \bigcup_{ptype\in PType}
      ptype$ is a \emph{total function} which gives the current value
      of each parameter.
    \end{itemize}
  \end{itemize}
\end{definition}

\end{document}